\documentclass[a4paper]{article}
\usepackage{graphicx}
\newcommand{\overset}[2]{ \mathop#2\limits^{#1} }

\begin{document}
\title{Interaction of axions with relativistic spinning particles}

\author{V A Popov and A B Balakin}

\date{
\small
Institute of Physics, Kazan Federal University, Kremlevskaya st. 18, Kazan 420008, Russia
}
\maketitle

%

\begin{abstract}
We consider a covariant phenomenological model, which describes an interaction between a pseudoscalar (axion) field and massive spinning particles. The model extends the Bagrmann-Michel-Telegdy approach in application to the axion electrodynamics. We present some exact solutions and discuss them in the context of experimental tests of the model and axion detection.
\end{abstract}

\section{Introduction}

Massive pseudo-Goldstone bosons were postulated in 1977 by Peccei and Quinn \cite{PQ} in the context of the strong $CP$-invariance problem, and were introduced into the high-energy physics as new light bosons or axions by Weinberg \cite{Weinberg} and Wilczek \cite{Wilczek} in 1978. Later axions with masses in the range of $10^{-6}-10^{-3}$ eV have acquired the status of one of the dark matter candidates \cite{A1,A2}, and were included into the experimental "short list" among other axion-like particles (ALPs) \cite{ALPs1,ALPs2}.

One of the trends in the experimental detection of axions use the idea of the axion-photon conversion, which is described mathematically by the term $\phi\, F_{ij}\overset{*}{F}{}^{ij}$ introduced into  the Lagrangian of the model \cite{WTN77,Sikivie}. Here $\phi$ is the pseudoscalar (axion) field, $F_{ij}$ is the Maxwell tensor and $\overset{*}{F}{}^{ij}$ is its dual. The alternative trend in the experimental activity is based on the idea of influence of the axions on nuclear spins, electric and magnetic moments, etc. (see, e.g., \cite{E1,E5,E6,E7,E8,E9}). Mathematically, such interactions can be described by the term $\overline{\psi}\gamma^5\gamma^i\psi\nabla_i\phi$, related to the axion-fermion coupling; here $\psi$ is the Dirac field attributed to fermions. This branch of axion physics can be indicated as spin-axion coupling by analogy with the axion-photon and the axion-gluon couplings. Nevertheless the term \emph{spin-axion coupling} is rather wide and two theoretical approaches have to be distinguished. Experimental equipments in the terrestrial laboratories deal with \emph{bonded} spin particles in the specific material media, and these bonded particles are, in average, at rest with respect to these media. The non-relativistic formalism of condensed matter physics is adequate for these purposes and it correctly describes spin dynamics in this framework.

When one deals with \emph{free} polarized particles, which move with high velocity (e.g., in storage rings) or in a vicinity of sources of strong gravitational and axion fields (e.g., near axion stars), the covariant formalism of high energy physics is necessary for description of the model. The appropriate phenomenological approach was developed  in \cite{BP2015}. It generalizes the Bagrmann-Michel-Telegdy (BMT) model \cite{BMT} for particles with spin by introducing force-like terms, which include the pseudoscalar field and its four-gradient, the Riemann curvature tensor in convolutions with the Maxwell tensor and its dual.

In this note we attract the attention to effects, which can accompany the motion of a relativistic particle with spin through the axion dark matter surrounding the Earth.

\section{Generalized BMT model}

In order to describe the behavior of charged relativistic point particles with a momentum four-vector $p^i$ and a spin four-vector $S^k$ we use the equations
\begin{equation}\label{eq : ParticleDyn}
\frac{D p^i}{D\tau} = {\cal F}^i , \quad \frac{D S^i}{D\tau} = {\cal G}^i .
\end{equation}
The scalar squares of the momentum $p^i$ and the spin $S^i$ four-vectors are assumed to be constants, i.e. $p^i p_i = m^2$ and $S^i S_i = - {\cal S}^2$, and also these vectors are orthogonal, i.e. $S^ip_i = 0$. These conditions provide the following general properties of the force-like quantities ${\cal F}^i$ and ${\cal G}^i$:
\begin{equation}\label{eq : GeneralProperties}
p_i{\cal F}^i = 0 , \quad S_i{\cal G}^i = 0 ,\quad {\cal F}^i  S_i + {\cal
G}_i p^i =0 ,
\end{equation}
which can be readily obtained from the equations of motion (\ref{eq : ParticleDyn}) convolving them with the momentum $p^i$ and the spin $S^i$, respectively. The equations (\ref{eq : GeneralProperties}) are
satisfied identically, when
\begin{equation}\label{eq : BMTapproach}
{\cal F}^i  = \omega^{ik} p_k  \,, \quad {\cal G}^i = \omega^{ik} S_k \,,
\end{equation}
where $\omega^{ik}$ is an arbitrary antisymmetric tensor, $\omega^{ik}= - \omega^{ki}$. Contents of this tensor predetermine particle dynamics. For the BMT model this tensor  is of the form (we use the system of units with $c=1$.)
\begin{equation}\label{eq : BMTomega}
\omega^{ik} = \omega^{ik}_{(0)} {=} \frac{e}{2m} \left[g F^{ik} +
\frac{\left(g{-}2 \right)}{m^2} \delta^{ik}_{mn} p_{j}F^{jm}p^n
\right] ,
\end{equation}
where $\delta^{ik}_{mn}=\delta^{i}_{m}\delta^{k}_{n}-\delta^{i}_{n}\delta^{k}_{m}$ is the four-indices Kronecker tensor, and $g$ is the dimensionless gyromagnetic ratio ($g$-factor).

To develop this approach with the axion field we introduce the tensor $\omega^{ik}$ linear in the pseudoscalar field $\phi$ or in its gradient four-vector $\nabla_i \phi$. The corresponding parts are
\begin{equation}\label{eq : omegaAx1}
\omega^{ik}_{(1)} {=} \frac{e \lambda}{2m} \phi \ \left[g_{A}
\overset{*}{F}{}^{ik} {+} \frac{\left(g_{A}{-}2 \right)}{m^2}
\delta^{ik}_{mn} p_{j}\overset{*}{F}{}^{jm}p^n \right],
\end{equation}
and
\begin{eqnarray}
\omega^{ik}_{(2)} &=&
\frac{e \mu}{2m} p^l \nabla_l \phi \biggl[g_{G} \overset{*}{F}{}^{ik}  + \frac{(g_{G}-2)}{m^2}\delta^{ik}_{mn} p_{j}\overset{*}{F}{}^{jm}p^n +
\omega_{23} \epsilon^{ikjn} p_n   p^{s}F_{js} \biggr] {+}
\label{eq : omegaAx2}
\\ &&
+ \frac{\nabla_l \phi}{m}\left[ \omega_{24} \epsilon^{ikmn} p_n
F_{m}^{\ \ l}  {+} \omega_{25} \epsilon^{ikln} p_n \right],
\nonumber
\end{eqnarray}
where $\epsilon^{klmn}$ is the Levi-Civita symbol with $\epsilon^{0123}{=}1$. Some terms in (\ref{eq : omegaAx1}) and (\ref{eq : omegaAx2}) appear as axionic counterparts of the classical BMT form (\ref{eq : BMTomega}). They are provided with their own phenomenological coupling constants $\lambda$, $\mu$ and axionic analogs of the $g$-factor. Other terms have no direct analogs, and we
use the constants $\omega_{23}$, $\omega_{24}$, etc., to indicate them.

In this work we focus on some features of the model, which can be a matter of concern in experimental physics.
With this in mind  we will consider a relativistic particle dynamics in the uniformly distributed dark matter axion field, which depends on time only, $\phi(t)$. Also we neglect  gravitational effects and deal, in fact, with the flat (Minkowski) spacetime.
In this case evolution of the electromagnetic and axion fields obey the equations \cite{Sikivie,NM4}
\begin{equation}\label{eq : AxED}
\dot{\textbf{E}}=-\dot\phi\,\textbf{B}\, ,\quad \ddot\phi+m_a^2\phi=-\frac{1}{\Psi_0^2}\,\textbf{EB}\,,
\end{equation}
where $m_a$ is the axion mass, the constant $\Psi_0$ is proportional to the axion decay constant $f_a$. The axion field is dimensionless in this representation. The dot denotes a derivative with respect to time.

\section{Direct spin-axion coupling}

As a first illustration of the spin-axion interaction we consider the direct effect, which takes place even when the electromagnetic field is absent, $F_{ik}=0$. In this case Eqs.~(\ref{eq : ParticleDyn}) take the simple form
\begin{equation}\label{eq : DSAIeqs}
\frac{d p^{i}}{d\tau} = 0 \,, \qquad \frac{d S^{i}}{d\tau} =
\frac{\omega_{25}}{m}  \epsilon^{ikln} S_k \nabla_l \phi\, p_n  \,.
\end{equation}
Clearly, the first equation in (\ref{eq : DSAIeqs}) describes the motion of free particle, and the second one is the equation of spin precession induced by axions.
Indeed, there is an evident analogy between the spin equation in (\ref{eq : DSAIeqs}) and the equation for the spin in an external magnetic field, which can be represented via its four-vector $B_i\,$. The right-hand side of this equation is equal to $(-e/m^2)\epsilon^{ikln}S_k B_l p_n\,$, and both the equations has the same form with the correspondence $m\omega_{25} \nabla_l \phi \rightarrow - e B_l\,$. Owing to this analogy it is clear that the gradient of the axion field can produce spin rotation similar to the well-known effect in the magnetic field. An important difference is that $B_k$ is a spacelike vector, while the gradient $\nabla_k \phi$ can be not only a spacelike, but also timelike or null vector. As a result, the magnetic field always rotates the spin except the case, when the three-momentum is parallel to the spin, whereas the axionically induced spin precession is a phenomenon much more sophisticated. In our case the precession appears only if the axion field is non-stationary, and the particle moves with respect to the axion environment.

Let the particle moves along the $x^3$-axis with the velocity $V=p^3/p^0$. In this case the $S^0$ and $S^3$ components of the spin four-vector remain constant and the precession occurs only in the transverse plane with respect to the direction of motion. Without loss of generality we can take the transversely polarized particle (i.e., particles with $S^0=S^3=0$) to obtain
\begin{equation}
S^1(t) = {\cal S} \cos \left[\omega_{25}V\phi(t)\right] \,,
\quad  S^2(t) = {\cal S} \sin \left[\omega_{25}V\phi(t)\right] \,.
\label{1A12}
\end{equation}
When the charged ultrarelativistic particle moves in a storage ring, the precession of this type also takes place (see \cite{BP2015} for details). If the polarized beam of electrons is formed in a storage ring, so that all the spins are initially directed perpendicularly to the ring plane, one can expect, that axions will rotate the polarization of the beam. In order to estimate the possible total axionically induced spin turn $\Delta S$, one can use the relation between the dimensionless pseudoscalar field and the energy density of the relic axion cold dark matter, given by  $(1/2)(m_a\Psi_0\phi)^2\approx \rho_{\rm{DM}}$. Taking into account that the local galactic halo dark matter energy density near the Earth is estimated as $\rho_{\rm{DM}}\sim 1$ GeV/cm$^3$  and $m_a\Psi_0\sim m_af_a\sim 10^{15}$ eV, we have
\begin{equation}
\Delta S\sim 10^{-17}\left(\frac{\omega_{25}}{1}\right)\left(\frac{\rho_{\rm{DM}}}{1\mbox{ GeV}/\mbox{cm}^3}\right)^{1/2}.
\end{equation}
Theoretically, the effect is measurable with modern high-precision instrumentation. Practically, it is very tiny to use the axionically induced spin precession of relativistic particles as an independent experiment.

\section{Axionic BMT-like coupling}

To illustrate the indirect spin-axion coupling, we consider the term $\omega^{ik}_{(1)}$ describing by Eq.~(\ref{eq : omegaAx1}) with only one non-vanishing phenomenological constant $\lambda\ne 0$, and $g=g_A=2$, for the case, when the electric and magnetic fields are directed along $x^3$-axis.
Now the equations for the particle motion (\ref{eq : ParticleDyn}) are of the form
\begin{equation}\label{eq : AxBMTdyn}
\frac{d p^{i}}{d\tau}=\frac{e}{m} \left( F^{ik} + \lambda \phi \overset{*}{F}{}^{ik} \right)p_k\,,\quad \frac{d S^{i}}{d\tau}=\frac{e}{m} \left( F^{ik} + \lambda \phi \overset{*}{F}{}^{ik} \right)S_k
\end{equation}
and have the BMT-like form for the particle without an anomalous magnetic moment.

The simple solution describing all main properties of the considered system is obtained if we take the following initial data: $p^3(0)=0$, $p^1(0)=0$, and $p^2(0)= q$. The transverse (with respect to the magnetic field) spin component is taken to be perpendicular to the three-momentum, i.e., $S^1(0)=\sigma$, $S^2(0)=0$.
In this case the solution to Eqs.~(\ref{eq : AxBMTdyn}) is
\begin{equation} \label{eq : Pi gen}
\begin{array}{ll}
p^0(t)=\left[m^2+q^2+\Phi(t)^2\right]^{1/2},
&\displaystyle S^0(t)= \sqrt{\frac{{\cal S}^2-\sigma^2}{m^2+q^2}} \,\Phi(t) ,\\
p^1(t)=q\sin \Psi(t),& S^1(t)=-\sigma\cos \Psi(t),\\
p^2(t)=q\cos \Psi(t),& S^2(t)=\sigma\sin \Psi(t),\\
p^3(t)=\Phi(t),
&\displaystyle S^3(t)=\sqrt{\frac{{\cal S}^2-\sigma^2}{m^2+q^2}}  \left[m^2+q^2+\Phi(t)^2\right]^{1/2},
\end{array}
\end{equation}
The time dependent functions $\Phi$ and $\Psi$ are defined by
\begin{equation}\label{eq : PhiPsi1}
\Phi(t)=\int\limits_{0}^t e(E_z+\lambda\phi B_z)dt'\,, \quad
\Psi(t)=\int\limits_{0}^t \frac{e}{p^0} (B_z-\lambda\phi E_z)dt'\,.
\end{equation}
When we deal with storage rings, the magnetic field $B_z$ can be taken as a constant. According to Eq.~(\ref{eq : AxED}) the axions generate the electric field $E_z(t)=-\phi(t) B_z$, and thus, the functions (\ref{eq : PhiPsi1}) can be rewritten as
\begin{equation}\label{eq : PhiPsi2}
\Phi(t)=e B_z(\lambda-1)\int\limits_{0}^t \phi (t')dt'\,, \quad
\Psi(t)=eB_z\int\limits_{0}^t \frac{1+\lambda\phi^2(t')}{\left[m^2+q^2+\Phi(t')^2\right]^{1/2}}dt'\,.
\end{equation}
When the relic cosmological axions are absent, the electric field vanishes, and $\Phi(t)=0$. Clearly, in this case (\ref{eq : PhiPsi2}) recovers the well-known solution describing the circular particle motion in the $(x^1,x^2)$-plane with the Larmor frequency $\Omega_L=eB_z/(m^2+q^2)^{1/2}$. The longitudinal spin component does not changed while the transverse polarization precess so its direction is preserved with respect to the particle frame.

In the axion environment the particle  moves along a trajectory much more sophisticated. The electric field gives a momentum along the $x^3$-axis to the particle, which start to oscillate near the rotation plane. The rotation phase depends now on a vertical position of the particle. Such behavior is typical for any value of the coupling constant $\lambda$, including $\lambda=0$, when the corresponding interaction is lacking and the axions effect on the particle through the Lorentz force.

The oscillation frequency is equal to the frequency of the axion field $m_a$, therefore a maximal deflection $d$ from the rotation plane during the one period can be determined from the last line of (\ref{eq : Pi gen}) as $d\sim \Omega_L\phi/m_a^2$. For the axion cold dark matter one finds
\begin{equation}\label{eq : estim deflect}
d\sim 10^{-15}{\rm{ cm }}\left(\frac{\rho_{\rm{DM}}}{1\mbox{ GeV}/\mbox{cm}^3}\right)^{1/2} \left(\frac{ 10^{-5} {\rm{ eV}} }{{{m_a}}}\right)^2.
\end{equation}
There is a special case, when $\lambda=1$, so that the interaction term suppresses the Lorentz force and the particle moves round a circle in the $(x^1,x^2)$-plane without a vertical deflection. The rotation phase is ''almost'' Larmor
\begin{equation}
\Psi(t)=\Omega_L \left(t+\int\limits_{0}^t \phi^2 dt'\right)\,,
\end{equation}
and keeping in mind that $\phi$ is a small quantity we see that this motion is very similar to particle motion without axions.
In other words, additional interactions can mask the standard axion effects. This means that experimental detection of axions can be more difficult problem, than it seems now.

The deflection (\ref{eq : estim deflect}) is most likely to be unobservable for ''heavy'' axions with masses more than $10^{-7}$ eV, but can be, in principle, detectable for ''light'' axions and ALPs. However, the fact that until  now we did not see this type of deflection, can indicate that channels of interaction via the axionic Lorentz-like force is activated by the axion environment.

\section{Conclusions}

At present the best accuracy is achieved in experiments based on axion-photon conversions.
Alternative methods are based on effects of polarization dynamics. We think that a hybrid spin-axion-photon methods for the detection of evidences of the axionic dark matter deserves attention of experimentalists. Now the experimental investigations are focused on the search for axions with masses in the range of $10^{-6}-10^{-3}$ eV. New methods suitable for masses below  $10^{-6}$ eV are also developed \cite{E9,Sikivie3,squid}.
Axion detectors, which use relativistic polarized particles, are not elaborated now, however, they could become a reasonable alternative in the nearest future.

\vspace{\baselineskip}

This work was supported by Program of Competitive Growth of KFU (Project No.~0615/06.15.02302.034), and by Russian Foundation for Basic Research (Grant RFBR No.~14-02-00598).


\bibliography{jpcsrefs}

\end{document}